\documentclass[12pt]{article}

\usepackage[breaklinks]{hyperref}  % comment out if you do not have it
\usepackage{graphicx}
\usepackage[active]{srcltx} % comment out if you do not have it
\usepackage{sariel,wide}
\usepackage{enumerate}
\usepackage{ifpdf}
\usepackage{mcolors2}
\usepackage{amstext}
\usepackage{picins}

\newcommand{\PntSet}{{\mathsf{Q}}}
\newcommand{\PntSetA}{{\mathsf{P}}}
\newcommand{\ObjSet}{\mathsf{S}}
\newcommand{\Graph}{\mathsf{G}}
\newcommand{\TriSet}{\EuScript{T}}

\providecommand{\ComplexityClass}[1]{{{\textcolor[named]{\si{OliveGreen}}{%
      \textsc{#1}}}}}
\providecommand{\APXHard}{{\ComplexityClass{\si{APX}-Hard}}\xspace}
\providecommand{\APXHardness}{{\ComplexityClass{\si{APX}-Hardness}}\xspace}

\providecommand{\MaxSNPHard}{{\ComplexityClass{Max{S}NP-Hard}}\xspace}

\providecommand{\NP}{\ComplexityClass{NP}\xspace}
\providecommand{\POLYT}{\ComplexityClass{P}\xspace}
\providecommand{\PTAS}{\textsf{\si{PTAS}}\xspace}
\newcommand{\pnt}{\mathsf{p}}

\newcommand{\Family}{\mathcal{F}}
\newcommand{\FamilyA}{\mathcal{H}}

\begin{document}

\title{Being Fat and Friendly is Not Enough}
%{A Note on Set-Cover by Fat Triangles}

\author{Sariel Har-Peled\SarielThanks{}}

\date{\today}

\maketitle

\begin{abstract}
    We show that there is no $(1+\eps)$-approximation algorithm for
    the problem of covering points in the plane by minimum number of
    fat triangles of similar size (with the minimum angle of the
    triangles being close to $45$ degrees). Here, the available
    triangles are prespecified in advance. Since a constant factor
    approximation algorithm is known for this problem
    \cite{cv-iaags-07}, this settles the approximability of this
    problem.

    We also investigate some related problems, including cover by
    friendly fat shapes, and independent set of triangles in three
    dimensions.
\end{abstract}

\section{Introduction}

In the \emphi{planar set-cover} problem, we are given a set $\PntSet$
of points in the plane, and a set of shapes $\ObjSet$, and we would
like to find a minimum cardinality subset of shapes of $\ObjSet$, such
that their union covers all the points of $\PntSet$.

In \secref{basic}, we show that this problem is \MaxSNPHard even when
the shapes are fat, convex, of the same size, and their union
complexity is linear. As such, no \PTAS for such a problem is possible
(unless $\POLYT = \NP$). We remind the reader that a \PTAS (Polynomial
Time Approximation Scheme) is an approximation algorithm that given an
input and a parameter $\eps>0$ it outputs a $(1+\eps)$-approximate
solution to the given instance in polynomial time.

In fact, the same result holds even if the shapes are fat triangles,
with minimum angle approaching $45$ degrees and of similar size. In
fact, in this case, all the triangles are rotated and translated
``noisy'' copies of the same triangle, where the original triangle is
right-angled and also isosceles, with the two base angles being $45$
degrees. (The word ``noisy'' here means that we can make all the
lengths of the edges of the copies of this triangle to be arbitrarily
close to the lengths of the corresponding original edges.)
    
Finally, in \secref{independent}, we show that there is no \PTAS for
independent set of triangles in $\Re^3$. Here, we are given a set of
triangles in three dimensions, and we are looking for the largest
subset of triangles such that no pair of them intersects.

\paragraph{Known results.}
For relevant results see \cite{aes-ssena-09, cch-smcpg-09,
   mr-irghsp-09} and references therein. In particular, our basic
construction of \secref{basic}, is similar to one of the hardness
proofs of \cite{el-dgig-08}. Several \APXHardness results are known in
geometry, among them hardness of (i) independent set of boxes in 3d
\cite{cc-ahopi-05}, (ii) maximizing guarded boundary in art gallery
problems \cite{fmz-mgbag-07}, and (iii) separating points by
axis-parallel lines \cite{cdkw-spapl-05}. This list is by no means
exhaustive.

%%%%%%%%%%%%%%%%%%%%%%%%%%%%%%%%%%%%%%%%%%%%%%%%%%%%%%%%%%%%%%%%%%
%%%%%%%%%%%%%%%%%%%%%%%%%%%%%%%%%%%%%%%%%%%%%%%%%%%%%%%%%%%%%%%%%%

\section{Hardness of approximation of the friendly
   geometric set-cover problem}
\seclab{basic}

Consider a set system $(U, \Family)$, where $\cardin{U} = n$, and
every element of $\Family$ is a subset of $U$ of size at most $k$,
where $k$ is some fixed constant. We are interested in finding the
minimum cardinality cover of $U$ by sets of $\Family$. This is known
as the \emphi{minimum $k$-set cover} problem and it is \MaxSNPHard for
$k \geq 3$ \cite{acgkm-ca-99}. It is known that if a problem is
\MaxSNPHard then there is no \PTAS for it unless $\POLYT=\NP$. Here,
one can even assume that every point of $U$ participates in at most
$k+1$ sets of $\Family$.

\begin{defn}
    Let $\PntSet$ be a set of $n$ points in the plane, and let
    $\Family$ be a set of $m$ regions in the plane, such that 
    \begin{enumerate}[(i)]
        \item the shapes of $\Family$ are convex, fat, and of similar
        size,

        \item the boundaries of any pair of shapes of $\Family$
        intersect in at most $6$ points,

        \item the union complexity of any $m$ shapes of $\Family$ is
        $O(m)$,

        \item and, any point of $\PntSet$ is covered by a constant number
        of shapes of $\Family$.
    \end{enumerate}
    We are interested in the problem of finding a minimum number of
    shapes of $\Family$ that covers all the points of $\PntSet$. We
    will refer to this variant as the \emphi{friendly geometric set
       cover} problem.
\end{defn}

\newcommand{\DiskOrg}{\mathsf{disk}}

\begin{lemma}
    There is no \PTAS for the friendly geometric set cover problem,
    unless $\POLYT=\NP$.
\end{lemma}

\begin{proof}
    We will reduce an instance $(U, \Family)$ of the minimum $k$-set
    cover problem (for $k=3$) into an instance of the friendly
    geometric set cover problem. So, let $U = \brc{u_1,\ldots, u_n}$,
    and $\Family = \brc{S_1, \ldots, S_m}$. We place $n$ points
    equally spaced on the unit radius circle centered at the origin,
    and let $\PntSet = \brc{\pnt_1, \ldots, \pnt_n}$ be the resulting
    set of points. Let $f( u_i) =\pnt_i$, for $i=1,\ldots,n$. Next, we
    map the set $S_i$ (which is of size at most $k$) to the region
    \[
    R_i = \CH\pth{ \; \DiskOrg\pth{ 1-\frac{i}{10n^2m}} \; \cup \;
       f(S_i) \;},
    \]
    for $i=1,\ldots,m$, where $\CH$ is the convex hull, $f(S_i) =
    \cup_{x \in S_i} \brc{ f(x)}$, and $\DiskOrg( r )$ denotes the disk of
    radius $r$ centered at the origin. Visually, $R_i$ is a disk with
    three (since $k=3$) teeth coming out of it, see
    \figref{intersection:of:regions}. Note, that the boundary of two
    such shapes intersects in at most $6$ points.

    It is now easy to verify that the resulting instance of geometric
    set cover $\pth[]{\PntSet, \brc{R_1, \ldots, R_m}}$ is in fact
    friendly, and clearly any cover of $\PntSet$ by these shapes can
    be interpreted as a cover of $U$ by the corresponding sets of
    $\Family$. Thus, a \PTAS for the friendly geometric set cover
    problem, would imply a \PTAS for the minimum $k$-set cover, which
    is impossible unless $\POLYT=\NP$.
\end{proof}

\begin{figure}
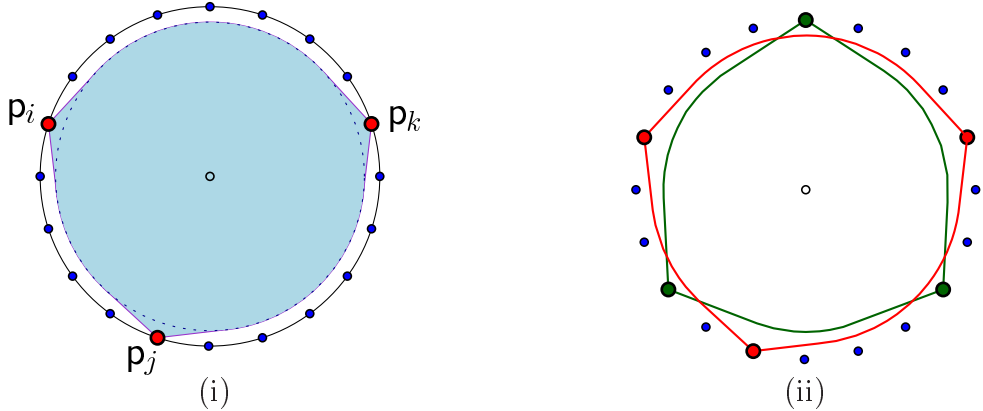

%    \centerline{\includegraphics{figs/gear}}
%    \newcommand{\IncGraphPage}[3]{
    \begin{tabular}{cc}
        \begin{minipage}{0.45\linewidth}
            \centerline{\IncGraphPage{figs}{gear}{1}}
        \end{minipage}&
        \begin{minipage}{0.45\linewidth}
            \centerline{\IncGraphPage{figs}{gear}{2}}
        \end{minipage}\\
        (i) & (ii)
    \end{tabular}

    \caption{(i) A region $R$ constructed for the set $S_t = \brc{u_i,
          u_j, u_k}$. Observe that in the construction, the inner disk
       is even bigger. As such, no two points are connected by an edge
       of the convex-hull when we add in the inner disk to the
       convex-hull. As such, each point ``contribution'' to the region
       $R$ is separated from the contribution of other points. (ii)
       How the intersection of the boundary of two such regions looks
       like.}

    \figlab{intersection:of:regions}
\end{figure}

\section{Hardness of approximation for set-cover 
   by fat triangles}

It is known that \ProblemC{Vertex Cover} is \APXHard even for a graph
with a maximum degree $3$ \cite{acgkm-ca-99}. A problem that is
\APXHard does not have a \PTAS unless $\POLYT = \NP$.  Consider such a
graph $\Graph$, and observe that a \ProblemC{Vertex Cover} problem in
such a graph can be reduced to \ProblemC{Set Cover} where every set is
of size at most $3$. Indeed, the ground set $U$ is the edges of
$\Graph$, and every vertex $v \in V(G)$ gives a rise to the set $S_v =
\brc{e \sep{ v\in e \text{ and } e\in E(G)}}$, which is of size at
most $3$. Clearly, any cover $C$ of size $t$ for the set system
$\mathcal{X} = \pth{ U, \brc{S_v \sep{v \in V(\Graph)}}}$, has a
corresponding vertex cover of $\Graph$ of the same size. Thus,
\ProblemC{Set Cover} with every set of size (at most) three is
\APXHard (this is of course well known). Note, that in this set cover
instance, every element participate in exactly two sets (i.e., the two
vertices adjacent to the original edge).

The graph $\Graph$ is of maximum degree three, and by Vizing's theorem
\cite{bm-gta-76}, it is $4$ edge-colorable\footnote{Vizing's theorem
   states that a graph with maximum degree $\Delta$ can be edge
   colored by $\Delta+1$ colors. In this specific case, one can reach
   the same conclusion directly from Brook's theorem.}.  Thus, the
ground set of the set system $\mathcal{X}$ can be colored by $4$
colors, and no set in this set system has a color appearing more than
once.

In the \emphi{fat-triangle set cover problem}, specified by a set of
points in the plane $\PntSet$ and a set of fat triangles $\TriSet$,
one wants to find the minimum subset of $\TriSet$ such that its union
covers all the points of $\PntSet$.

\begin{lemma}
    There is no \PTAS for the fat-triangle set cover problem, unless
    $\POLYT=\NP$. 

    In fact, one can prespecify an arbitrary constant $\delta > 0$, and
    the claim would hold true even if the following conditions hold on
    the given instance $(\PntSet,\TriSet)$:
    \begin{center}
    \begin{minipage}{0.9\linewidth}
    \begin{enumerate}[(A)]
        \item The minimum angle of all the triangles of $\TriSet$ is
        larger than $45-\delta$ degrees.

        \item No point of $\PntSet$ is covered by more than two
        triangles of $\TriSet$.

        \item The points of $\PntSet$ are in convex position.

        \item All the triangles of $\TriSet$ are of similar
        size. Specifically, all the triangles diameter is in the range
        (say) $(2-\delta, 2]$.

        \item Each triangle of $\TriSet$ has two angles in
        the range $(45-\delta, 45+\delta)$, and one angle in the
        range $(90-\delta, 90+\delta)$.

        \item The vertices of the triangles of $\TriSet$ are the
        points of $\PntSet$.
    \end{enumerate}
    \end{minipage}
    \end{center}

    \lemlab{no:p:t:a:s}
\end{lemma}

\begin{proof}
    We are given an instance of the vertex cover problem for a graph
    with maximum degree $3$, and we transform it into a set cover
    instance as mentioned above, denoted by $\mathcal{X} = \pth{ U,
       \Family_{\mathcal{X}}}$. Let $n = \cardin{U}$, and color $U$
    (as described above) by $4$ colors such that no set of
    $\mathcal{X}$ has the same color repeated twice, let $U_1, \ldots,
    U_4$ be the partition of $U$ by the color of the points.
    
    \parpic[r]{\includegraphics{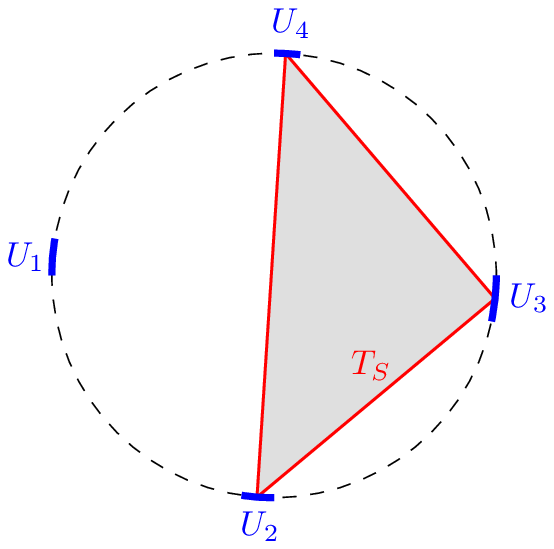}}
    
    Let $\mathcal{C}$ denote the circle of radius one centered at the
    origin. We place four relatively short arcs on $C$, placed on the
    four intersection points of $C$ with the $x$ and $y$ axises, see
    figure on the right.  Let $I_1, \ldots, I_4$ denote these four
    circular intervals.  We equally space the elements of $U_i$ (as
    points) on the interval $I_{i}$, for $i=1,\ldots, 4$. Let
    $\PntSet$ be the resulting set of points.

    For every set $S \in \Family_{\mathcal{X}}$, take the convex hull
    of the points corresponding to its elements as its representing
    triangle $T_S$. Note, that since the vertices of $T_S$ lie on
    three  intervals out of $I_1, I_2, I_3, I_4$, it follows
    that it must be fat, for all $S \in \Family_{\mathcal{X}}$. As
    such, the resulting set of triangles $\TriSet = \brc{T_S \sep{S
          \in \Family_{\mathcal{X}}}}$ is fat, and clearly there is a
    cover of $\PntSet$ by $t$ triangles of $\TriSet$ if and only if
    the original set cover problem has a cover of size $t$.

    In fact, any triangle having its three vertices on three different
    intervals of $I_1, \ldots, I_4$ is close to being an isosceles
    triangle with the middle angle being $90$ degrees. As such, by
    choosing these intervals to be sufficiently short, any triangle of
    $\TriSet$ would have a minimum degree larger than, say,
    $45-\delta$ degrees, and with diameter in the range between
    $2-\delta$ and $2$.

    This is clearly an instance the fat-triangle set cover
    problem. Solving it is equivalent to solving the original vertex
    cover problem, but since it is \APXHard, it follows that the
    fat-triangle set cover problem is \APXHard.
\end{proof}

\begin{remark}
    For fat triangles of similar size a constant factor approximation
    algorithm is known \cite{cv-iaags-07}.  \lemref{no:p:t:a:s}
    implies that one can do no better. Naturally, it might be possible
    to slightly improve the constant of approximation provided by the
    algorithm of Clarkson and Varadarajan \cite{cv-iaags-07}.

    However, for fat triangles of different sizes, only a $\log \log
    \log$ approximation is known \cite{aes-ssena-09}. It is natural to
    ask if this can be improved.
\end{remark}

\subsection{Extensions}

\begin{lemma}
    Given a set of points $\PntSet$ in the plane and a set of circles
    $\Family$, finding the minimum number of circles of $\Family$ that
    covers $\PntSet$ is \APXHard; that is, there is no \PTAS for this
    problem.

    \lemlab{no:P:T:A:S:circles}
\end{lemma}
\begin{proof}
    Slightly perturb the point set used in the proof of \lemref{no:p:t:a:s},
    so that no four points of it are co-circular. Let $\PntSet$
    denote the resulting set of points. For every set $S \in
    \Family_{\mathcal{X}}$, we now take the circle passing through the
    three corresponding points. Clearly, this results in a set of
    circles (that are almost identical, but yet all different), such
    that finding the minimum number of circles covering the set
    $\PntSet$ is equivalent to solving the original problem.
\end{proof}

\begin{lemma}
    Given a set of points $\PntSetA$ in $\Re^3$ and a set of planes
    $\Family$, finding the minimum number of planes of $\Family$ that
    covers $\PntSetA$ is \APXHard; that is, there is no \PTAS for this
    problem.
\end{lemma}
\begin{proof}
    Let $\PntSet$ be the point set and $\Family$ be the set of circles
    constructed in the proof of \lemref{no:P:T:A:S:circles}, and map
    every point in it to three dimensions using the mapping $f: (x,y)
    \rightarrow (x,y,x^2 + y^2)$.  This is a standard lifting map used
    in computing planar Delaunay triangulations via convex-hull in
    three dimensions, see \cite{bkos-cgaa-00}.  Let $\PntSetA =
    f(\PntSet)$ be the resulting point set.

    It is easy to verify that a circle of $c \in \Family$ is mapped by
    $f$ into a curve that lies on a plane. We will abuse notations
    slightly, and use $f(c)$ to denote this plane.  Let $\FamilyA =
    f(\Family)$. Furthermore, for a circle $c \in \Family$, we have
    that $f(c \cap \PntSet) = f(c) \cap \PntSetA$.  Namely, solving
    the set cover problem $(\PntSetA, \FamilyA)$ is equivalent to
    solving the original set cover instance $(\PntSet, \Family)$.
\end{proof}

\bigskip

Interestingly, the recent work of Mustafa and Ray \cite{mr-irghsp-09}
implies that there is a \PTAS for set cover of points by disks (i.e.,
circles with their interior), and similarly, there is a \PTAS for the
problem of set cover of points by half-spaces in three
dimensions. Thus, somewhat surprisingly, the ``shelled'' version of
these problems are harder than the filled-in version.

\section{Hardness of independent set of triangles in 3d}
\seclab{independent}

Given a set $\ObjSet$ of $n$ objects in $\Re^d$ (say, triangles in
3d), we are interested in computing a maximum number of objects that
are \emphi{independent}; that is, no pair of objects in this set
(i.e., independent set) intersects. This is the geometric realization
of the \emphi{independent set} problem for the intersection graph
induced by these objects.

\begin{lemma}
    There is no \PTAS for the maximum independent set of triangles in
    $\Re^3$, unless $\POLYT=\NP$.
\end{lemma}

\begin{proof}
    Independent set is \APXHard even for graphs with maximum degree
    $3$ \cite{acgkm-ca-99}. So, let $\Graph=(V,E)$ be a given such
    graph with maximum degree $3$, where $V = \brc{v_1,\ldots,
       v_n}$. We will create a set of triangles, such that their
    intersection graph is $\Graph$.

    We use the following fact: If one spreads $n$ points
    $\pnt_1,\ldots, \pnt_n$ on the positive branch of the moment curve
    in $\Re^3$ \cite{s-eubnf-91, ek-alnfc-03}, their Voronoi diagram
    is \emphi{neighborly}; that is, every Voronoi cell is a convex
    polytope that shares a non-empty two dimensional boundary face
    with each of the other cells of the diagram. Let $C_i$ denote the
    cell of the point $\pnt_i$ in this Voronoi diagram, for
    $i=1,\ldots, n$.

    Now, for every vertex $v_i \in V$, we form a set $\PntSet_i$ of
    (at most) three points, as follows. If $v_iv_j \in E$, then we
    place a point $p_{ij}$ on the common boundary of $C_i$ and $C_j$,
    and we add this point to $\PntSet_i$ and $\PntSet_j$.
    
    For $i=1,\ldots, n$, the region $f_i$ corresponding to $v_i$ is
    the triangle formed by the convex-hull of $\PntSet_i$ (if
    $\PntSet_i$ have fewer than three points then the triangle is
    degenerate).

    Let $\TriSet = \brc{f_1,\ldots, f_n}$.  Observe that the triangles
    of $\TriSet$ are disjoint except maybe in their common vertices,
    as their interior is contained inside the interior of $C_i$, and
    the cells $C_1, \ldots, C_n$ are interior disjoint. Clearly $f_i
    \cap f_j \ne \emptyset$ if and only if $v_iv_j \in E$. Thus,
    finding an independent set in $G$ is equivalent to finding an
    independent set of triangles of the same size in $\TriSet$. We
    conclude that the problem of finding maximum independent set of
    triangles is \APXHard, and as such does not have a \PTAS unless
    $\POLYT=\NP$.
\end{proof}

\bigskip

The above result is still a far cry from being tight. In light of the
results known \cite{am-isigc-06} for independent set of segments in
the plane, it is natural to conjecture that this harder problem (i.e.,
finding independent set of triangle) can not be approximated to within
any polynomial factor.

Implicit in the above proof is the fact that any graph can be realized
as the intersection graph of convex bodies in $\Re^3$ (we were a bit
more elaborate for the sake of completeness and since we needed
slightly more structure). This is well known and can be traced to a
result of Tietze from 1905 \cite{t-upnr-05}.

%------------------------------------------------------------------
%------------------------------------------------------------------

\section*{Acknowledgments}

The author thanks Janos Pach and Shakhar Smorodinsky for useful
discussions on this problem.

%-------------------------------------------------------------------------
 
\bibliographystyle{salpha} 
\bibliography{shortcuts,geometry}

\end{document}